# ADAPTIVE LOCOMOTION OF ARTIFICIAL MICROSWIMMERS

H.-W. Huang[1*], F. E. Uslu[2], P. Katsamba[3], E. Lauga[3], M. S. Sakar[2*], B. J. Nelson[1]†

1. Department of Mechanical Engineering, ETH Zurich, CH-8092, Zurich, Switzerland
2. Institute of Mechanical Engineering, Ecole Polytechnique Fédérale de Lausanne, CH-1015, Lausanne, Switzerland
3. Department of Applied Mathematics and Theoretical Physics, University of Cambridge, Cambridge CB3 0WA, United Kingdom

* H.-W Huang and M. S. S. contributed equally to this work
† To whom correspondence should be addressed. E-mail:bnelson@ethz.ch

**Abstract**

Bacteria can exploit mechanics to display remarkable plasticity in response to locally changing physical and chemical conditions. Compliant structures play a striking role in their taxis behavior, specifically for navigation inside complex and structured environments. Bioinspired mechanisms with rationally designed architectures capable of large, nonlinear deformation present opportunities for introducing autonomy into engineered small-scale devices. This work analyzes the effect of hydrodynamic forces and rheology of local surroundings on swimming at low Reynolds number, identifies the challenges and benefits of utilizing elastohydrodynamic coupling in locomotion, and further develops a suite of machinery for building untethered microrobots with self-regulated mobility. We demonstrate that coupling the structural and magnetic properties of artificial microswimmers with the dynamic properties of the fluid leads to adaptive locomotion in the absence of on-board sensors.

**Introduction**

Microorganisms manifest a diverse set of molecular motility machinery to effectively navigate complex environments and occupy a variety of ecological niches *(1)*. Swimming in bacteria arises from the mechanical interactions between the actuated flagella, cell body and the drag generated by the flow *(2, 3)*. Hydrodynamic drag is dominated by viscous forces at low Reynolds number which in turn depend on the shape of the moving object. Bacteria can adopt alternate shapes and sizes over the course of their life cycles to optimize their motility *(4–6)*. In addition to modulating cell body shape, bacteria can also utilize the form and structure of the propulsive system for advanced maneuverability in complex environments. Bending of the hook enhances motility in *Caulobacter crescentus (7)* while monotrichous *Vibrio alginolyticus* out-performs multiflagellated *Esherichia coli* in climbing nutrient gradients with the aid of a flagellar buckling instability *(8)*. A polymorphic transition in the flagellar filament enables *Shewanella putrefaciens* to escape from physical traps *(9)*.

The development of microscopic artificial swimmers that can cross biological barriers, move through bodily fluids, and access remote pathological sites can revolutionize targeted therapies *(10–13)*. Seminal work demonstrated the feasibility of following the example of prokaryotic *(14, 15)* or eukaryotic *(16)* flagellum for building magnetically controlled microswimmers that have the ability to exhibit non-reciprocal motion. However, unlike living cells, these mechanical devices neither sense their local environment nor react to changes in physical conditions. Addressing these issues with traditional robotic solutions based on electronic circuitry would require highly sophisticated manufacturing processes and result in orders of magnitude increase in the size of the machines. Utilization of biological actuators and sensors for engineering autonomous biohybrid robotic devices is an intriguing alternative *(17)*. Although the field is in its infancy, proof-of-concept examples have already demonstrated the potential *(18–20)*. Here, we focus on artificial materials to pave the way for building robust, tunable and durable engineering solutions.

Fluid-structure coupling in hydrogel-based compliant machinery may present a possible mechanism for autonomous regulation of morphology and function. Using origami design principles as a framework, a variety of folding techniques have been introduced for the development of three-dimensional (3D) flexible microstructures *(21)*. Production of programmable self-folding films at microscale can be achieved via patterning of multiple layers with different swelling properties *(22–24)* or creation of spatial concentration gradients *(25, 26)*. However, these methods provide limited control over the mechanical and magnetic properties of the machine. In our previous work, we have shown that the form and magnetization profile of self-folded micromachines can be independently programmed by incorporating magnetic nanoparticles into sequentially patterned hydrogel layers *(27)*. In this work, we introduce a simple and versatile method for engineering magnetically controlled soft micromachines as 3D reconfigurable multibody systems from a nanocomposite hydrogel monolayer. We present a set of design strategies for self-regulation of motility and maneuverability by utilizing the interplay among viscous, elastic, magnetic and osmotic forces. We show that reconfigurable body can continuously morph in accordance with the properties of the surrounding fluid, a feature that leads to passing through constrictions and enhancement of locomotion performance. Employing elastohydrodynamic coupling in shape-shifting and gait adaptation enables enticing opportunities for microrobots navigating inside obstructed, heterogeneous, and dynamically changing environments.

## Results and Discussion

**Building soft microswimmers with bioinspired locomotion.** We followed a variant of Origami, called Kirigami, to design and fold compliant 3D microstructures from a thermoresponsive gel composite reinforced with magnetic nanoparticles (MNPs). The fabrication process involves cutting initiated by photolithography and folding upon hydration of the polymerized layer. We generated nonuniform distribution of MNPs along the thickness direction to form two distinct layers of hydrogels with significantly different swelling ratios through sedimentation or application of magnetic forces. Differential swelling upon hydration along the film thickness resulted in self-folding of monolayer structures (see *S1. Programmable folding of monolayer nanocomposites*). The curvature of folded sheet was proportional to the MNP concentration (Fig. S1). The folding axis of each compartment was parallel to the alignment of encapsulated MNPs due to constrained swelling in the direction of reinforcement. Particle alignment was performed by the application of uniform magnetic fields during sample preparation (Fig. S2). The folding axis and the magnetic anisotropy of each compartment were coupled to each other and both defined by the orientation of the reinforcing MNPs. Hundreds of micromachines with complex 3D architectures were fabricated from the same film using a single-step photolithography process (Fig. 1*A*).

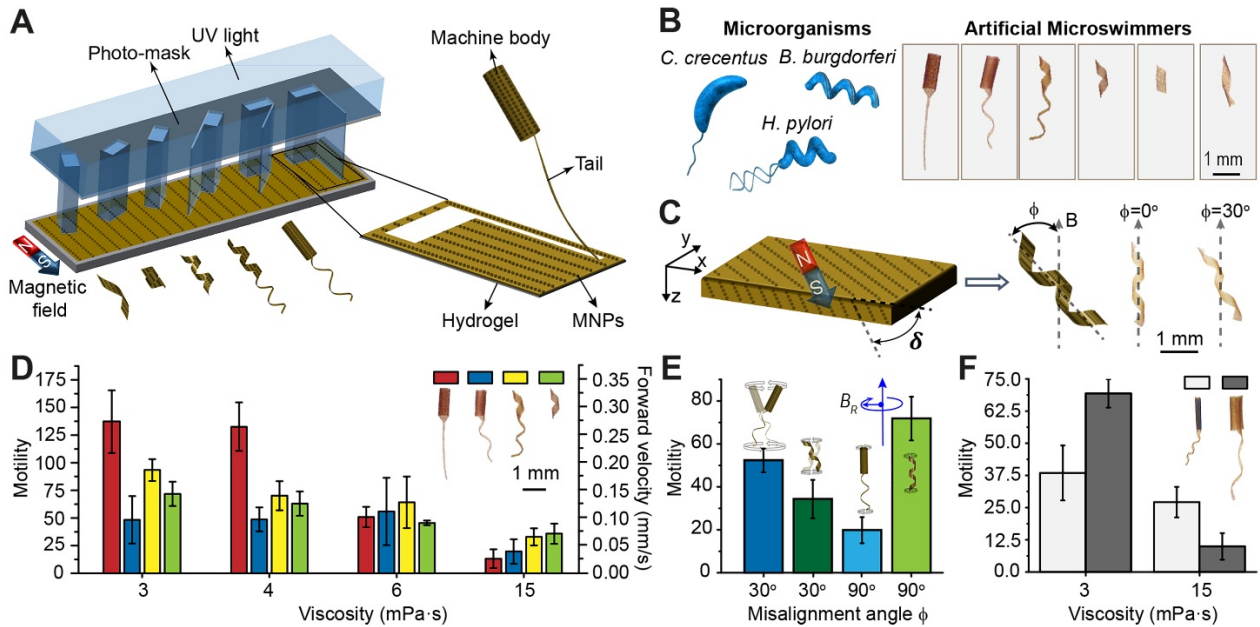

**Figure 1. Design, development, and actuation of bioinspired microswimmers.** (*A*) A Kirigami approach to build mass customized soft microswimmers through a single-step photolithography. (*B*) The schematic illustration of the bacteria taken as inspiration for this study and the optical images of the engineered artificial microswimmers. (*C*) Out of plane alignment ($\delta \neq 0$) of MNPs lead to non-zero misalignment angle $\phi$. The optical images showing two swimmers with identical shapes and varying $\phi$. (*D*) A comparison of the motility of microswimmers swimming in fluids with different viscosities. (*E*) The motility of the flagellated tubular microswimmers and helical microswimmers encoded with two different magnetic anisotropy rotating in a solution with a viscosity of 3 mPa·s. (*F*) The effect of body size on the motility of the tubular microswimmers. The swimmers were driven at 2 Hz with a field strength of 20 mT in all experiments, unless stated otherwise. All bar graphs represent average $\pm$ s.e.m. (N = 6 measurements for each microswimmer and three different swimmers tested per condition.

We focused on three different configurations inspired by widely studied microorganisms, *Caulobacter crescentus*, *Helicobacter pylori*, and *Borrelia burgdorferi* (Fig. 1*B*). Bacteria swim by rotating propeller-like organelles, called flagellar filaments, which extend from the cell body *(28)*. Artificial microswimmers can mimic this motion if the magnetic moment of the machine is perpendicular to its long axis *(29)*. We explored the effect of shape anisotropy on the magnetization profile of the microswimmers. In-plane (x-y plane) alignment of MNPs in the unfolded monolayer resulted in a magnetization parallel to the folding axis and a misalignment angle ($\phi$), which is the angle between the folding axis and magnetic fields, of zero. In other words, the structures resemble compass needles that align their long axis to the direction of the external magnetic field *(30)*. To address this limitation, we fabricated microswimmers with varying out-of-plane particle alignment while keeping in-plane particle alignment constant. Out-of-plane alignment of MNPs has no effect on the final 3D shape due to the relatively small thickness (~30 μm) of the monolayer compared to its overall size. With the magnetization component in z axis, folded structures acquired a magnetic moment in the radial direction, where $\phi$ was equal to the angle of the out-of-plane alignment of MNPs with respect to the x-y plane (Fig. 1*C*).

**Optimal motility at different viscosities requires different gaits.** Previous work on *C. crescentus* has shown that the flexibility of the hook generates cell body precession that leads to a 3D helical motion *(7)*. The slantwise motion of the cell body during precession develops thrust, adding to that developed by the flagellar filament. On the other hand, the helical shape of *Vibrio cholerae* enhances motility within a polymer network, a feature suggested to be important for its pathogenicity *(31)*. We systematically explored the potential advantage of this morphological diversity by building microswimmers with different body plans and actuating them in fluids with varying viscosity. Although we did not design a separate flexible hook that connects the tail and the body, we can still engineer microswimmers that follow 3D helical trajectories by coordinating their morphology with magnetization profile *(32)*. The Reynolds number is ranging from $10^{-2}$ to $10^{-4}$ in all the experiments presented in this work, thus swimming is performed under laminar flow. The normalized velocity of microswimmers is reported to provide a more accurate comparison of performance *(16)*. For this reason, we express motility as $U/fL * 1000$, where $U$ is the forward velocity, $f$ is the rotating frequency and $L$ is the body length of the micromachines *(33)*.

In a sucrose solution with a similar viscosity to blood (3 $mPa \cdot s$), the flagellated microswimmers with a tubular body and flexible planar tail moved much faster compared to other prototypes (Fig. 1*D*). The superior performance of this configuration compared to tubular body-helical tail and helical body-helical tail combinations can be explained by the enhanced body precession induced by the oar-like propulsion of the tail. Misalignment of the body with respect to the external magnetic field together with the flexibility of the tail lead to helical motion. The forward motility of the flagellated microswimmers was significantly higher when ϕ = 30° (Fig. 1*E*). While flagellated microswimmers benefited from both helical motion and corkscrew motion in this configuration, helical microswimmers suffered from extra drag generated by wobbling. Tuning ϕ to 90° resulted in a completely different picture. The motility of the flagellated swimmers dramatically dropped due to the absence of helical motion while helical microswimmers performed corkscrew motion without wobbling. These results show that the presence of a non-helical body is advantageous if the body can generate large amplitude helical motion.

The increase of viscosity monotonically decreased the motility of all microswimmers but the drop was drastic for flagellated microswimmers with a planar tail. With increasing viscosity, viscous forces start to attenuate helical motion of the cell body and the body becomes a source of passive drag that primarily impedes the motility. Furthermore, the lack of wobbling on the body eliminates bending of the tail and, in the absence of chirality, the tail cannot break the time-reversal symmetry to generate propulsion. Helical microswimmers were the fastest at high viscosity because the only relevant motion became the corkscrew motion. The body of microswimmers had a higher magnetic torque compared to their tail and thrust was mainly generated by the body, which made the tail obsolete at high viscosity. A larger body provided higher motility at low viscosity because those machines traced a helical path of higher amplitude (see *S2. Additional Discussion* and Fig. S3). However, at high viscosity, body precession was attenuated, and therefore, smaller body provided higher motility (Fig. 1*F*). Our results conflict with the argument that helical body does not provide a significant enhancement of motility for *H. pylori* inside viscous solutions *(34)*. The reason for this discrepancy lies in the body rotation; only the flagellum is actively rotated in bacteria and their bodies show a very slow counter-rotation to balance torque while we are simultaneously rotating the body and the tail with the same angular speed.

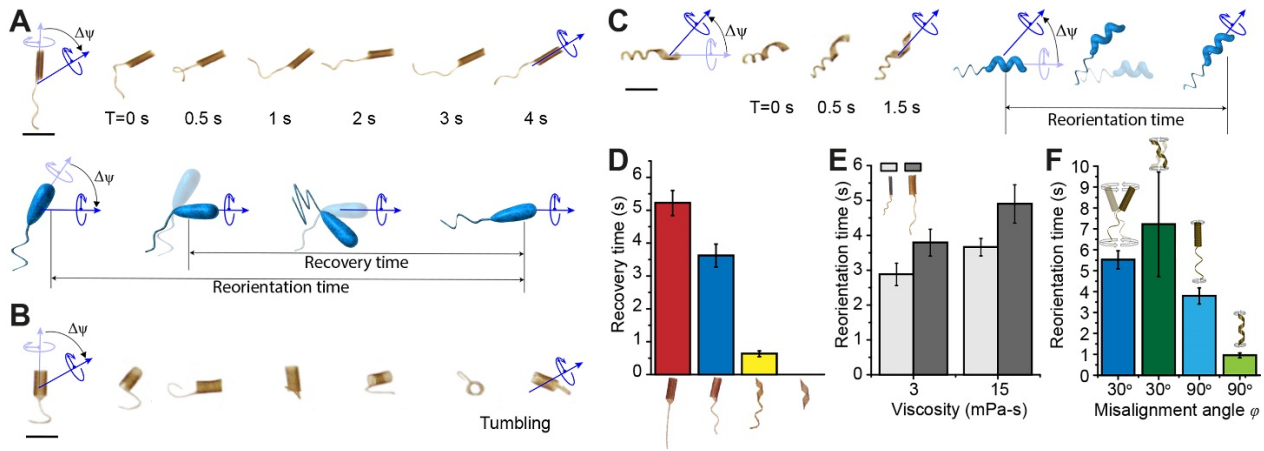

**Figure 2. Elastic instabilities and optimization of maneuverability.** The yaw angle Ψ was instantly changed 45° while the swimmers were driven were driven at 2 Hz with a field strength of 20 mT in a solution with a viscosity of 3 mPa·s. (*A*) Time-lapse images of a microswimmer with a tubular body and helical tail during the reorientation of its swimming direction. (*B*) Changing the tail to a planar geometry and ϕ to 30° led to complete loss of motility. (*C*) Changing the body geometry to a helix significantly reduces the reorientation time by providing instant recovery. (*D*) Quantitative comparison of reorientation time for different prototypes. (*E*) The effect of body size on the maneuverability of microswimmers swimming at varying viscosities. (*F*) The role of magnetic anisotropy on the maneuverability of the microswimmers with tubular and helical bodies. The scale bars are 1 mm. All bar graphs represent average ±s.e.m. (N = 6 measurements for each microswimmer and three different swimmers tested per condition.

**Morphology and magnetization profile together determine performance during navigation.** Along with motility, maneuverability plays a key role for bacteria in rapidly detecting and tracking nutrient gradients. By adjusting the relative frequency and length of reorientation phases, cells are able to adapt to the changing local chemical environment. Experimental evolutionary analysis of motile behavior of *E.coli* has shown that evolved strains had increased swimming velocity and frequency of reorientation, which together led to enhanced chemotaxis in porous media *(35)*. Experiments with *E.coli* have also shown that body size and shape control the average reorientation angle and time for a motile cell to change its direction of motion *(36)*. We investigated the maneuverability of artificial microswimmers by inducing deflections in the yaw angle during swimming (Fig. 2). A highly maneuverable microswimmer is expected to quickly change its movement direction with a small change in control signal. For gentle disturbances where the orientation of the rotating magnetic field is instantaneously changed for less than 10° around the yaw axis, all swimmers corrected their heading almost immediately. For stronger perturbations with 45° yaw rotation, both the body and tail geometry played an important role in the dynamic response of the compliant swimmers.

During a successful maneuver, the body responds to the control signal before the tail as the magnetization of the body is significantly higher. The speed of body rotation in response to the applied torque depends on the magnetization of the body and the hydrodynamic drag, which in turn are determined by the body geometry and magnetic volume. Interestingly, the rapid change in body orientation generated a buckling instability on the flexible tail for swimmers with tubular body. This instability led to a transient wobbling motion on the machine body until the tail reoriented with the main axis of the body, and this delay is quantified by the recovery time $\Delta t_{rec}$ (Fig. 2*A*). The overall delay between the change in the control signal and the completion of reorientation of the swimming direction is denoted by $\Delta t_{orient}$. Swimmers with a planar tail are more susceptible to instabilities. The contribution of helical tail to stabilization can be explained by the effectively higher stiffness of helical geometry compared to a planar structure and the attenuated precession on the body. At extreme cases ($\phi$ close to 0), change in direction resulted in tumbling motion, which manifests itself as loss of motility (Fig. 2*B*). Although inertial forces play no role at small scale, elastic instabilities occur due to the extreme compliance of the propulsion apparatus. One of the best demonstration in nature is the flicks in *V. alginolyticus*, which arises from an off-axis deformation of the flagellum caused by the buckling of the hook *(37)*. Switching the body morphology from a tube to a helix resulted in superior performance. The helical body generates propulsion together with the helical tail while applying pulling force on the filament, thus preventing buckling instability. As a result, the body and tail simultaneously reorient along the direction of magnetic field during maneuvers (Fig. 2*C*). While $\Delta t_{rec}$ is more than four seconds for the swimmers with a tubular body, it is less than 0.5 sec for swimmers with a helical body moving at the same velocity. Helical microswimmers showed the best performance as expected because they do not deal with body and tail coordination (Fig. 2*D* and Movie S1). We then explored the combined effect of body size and viscosity on maneuverability using the swimmers shown in Fig. 1*F*. Tubular machines with smaller folding diameter were prepared by increasing the nanoparticle concentration in the film formulation. Regardless of the value of viscosity, smaller body provided a comparative advantage by lowering rotational drag and increasing magnetic torque (Fig. 2*E* and see S2 for additional information).

We next studied how magnetization profile may affect maneuverability of microswimmers. So far, they were encoded with $\phi = 90°$ to prevent wobbling motion during forward swimming. Experiments on helical swimmers encoded with $\phi = 30°$ showed that wobbling motion can significantly affect $\Delta t_{orient}$ during maneuvers (Fig. 2*F*). Likewise, $\Delta t_{orient}$ of flagellated swimmers with $\phi = 30°$ was significantly higher than swimmers with $\phi = 90°$ because the rapid change in the yaw angle destabilized the machines and transformed the wobbling motion into a tumbling motion. To our surprise, adding a tail to the wobbling helical swimmer significantly enhanced its performance by completely preventing the occurrence of tumbling motion (Movie S2). These results pose a trade-off between motility and maneuverability at low viscosity as the reorientation time increases with decreasing $\phi$ and the motility increases with increasing $\phi$. The capability of dynamically re-magnetizing the body would provide a method to adjust the motility and maneuverability on demand. Magnetically reinforced nanocomposite were re-magnetized in a direction other than the direction of MNP alignment when the applied magnetic field was significantly higher than the magnetic field applied for the alignment of particles during fabrication (see *S3. Characterization of Magnetization Profile*, Fig. S4, and Movie S3).

**Exploiting elastohydrodynamic coupling for gait adaptation.** Unlike swimmers with rigid propulsion mechanisms, the coupling between magnetic forces, filament flexibility and viscous drag determines propulsion efficiency of compliant swimmers *(16)*. This nonlinear relationship is described by dimensionless Sperm number defined as

$$Sp = L_f / \left(\frac{A}{\xi_\perp \omega}\right)^{1/4}, \tag{1}$$

where $L_f$ is length of the flagellum, $A$ is the bending stiffness. For a slender filament ($L_f \gg a$) the perpendicular viscous coefficient is given by $\xi_\perp = \frac{4\pi\mu_d}{\log\left(\frac{L_f}{a}\right)+1/2}$. The radius $a$ is approximated by the geometric mean ($a = \sqrt{t \times w}$) of the thickness $t$ and width $w$ of the filament. For $Sp \ll 1$, bending forces dominate and the filament is effectively straight. Artificial microswimmers must operate away from this regime and optimal motility is predicted for $Sp$ of the order of unity.

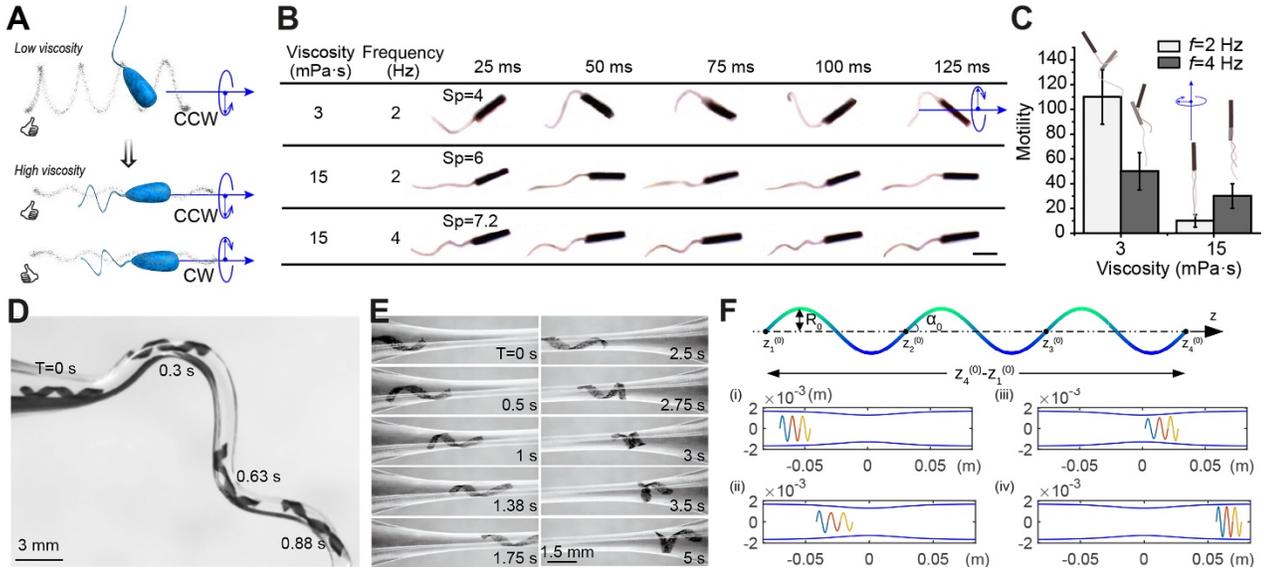

**Figure 3. The efficiency and mode of motility are controlled by the body plan.** (*A*) A schematic illustration of microswimmers swimming with an oar-like propulsion strategy at low viscosity and performing corkscrew motion at high viscosity due to coiling of the flexible tail. (B) The optical images and (*C*) motility along with schematic representations of microswimmers with shape-shifting tails moving at varying viscosities and rotating frequencies. The scale bar is 500 μm. (*D*) The optical images of a flexible helix passing through curved conduit morphologies with the flow rate of 2 ml/min. (*E*) Shape change driven by velocity gradients in a conduit with a constriction and the flow rate of 5 ml/min. (*F*) Computational model exploring shear-induced elongation in a conduit with a slowly varying constriction.

We asked whether the elastohydrodynamic properties can be exploited to trigger a gait transition in response to changes in viscosity (Fig. 3*A* and Movie S4). Analytical solution of the equations of motion for an actuated flexible tail predicted that the number of helical turns would increase with increasing $Sp$ (see *S4. Analytical solution for an actuated elastic filament* and Fig. S5). The bending stiffness was obtained from the measured values of the elastic modulus of the hydrogel and the filament geometry (Fig. S7). Fig. 3*B* shows time-lapse optical images of a tubular microswimmer with planar tail, encoded with ϕ = 30° swimming at different viscosities and rotating frequencies. As described before, this configuration generated a strong helical motion at low viscosity (3 mPa·s) that completely ceased at high viscosity (15 mPa·s). At relatively low viscosity and rotation frequency (*Sp* = 4), the internal and external stresses were mostly dissipated at the joint, which led to body precession. Constraining the body precession by setting ϕ to 90° reduced motility and enabled coiling in the tail. Coiling of the tail was clearly observable at higher viscosity and frequency (*Sp* = 7.2) both in the analytical solution (Fig. S5) and experimental data (Fig. 3*B*). This morphological transformation led to emergence of corkscrew motion and enhanced motility (Fig. 3C and Fig. S5D).

**Shape adaptation in complex channels under viscous flow.** Gradients in ambient fluid velocity is pervasive in microbial habitats and bacteria exhibit directed movement responses due to shear by utilizing their body shape *(38, 39)*. The extraordinary flexibility of red blood cells enables them to change shape under shear forces as they pass through vessels significantly smaller than their diameter *(40)*. Inspired by these elastohydrodynamic features, we exposed helical microswimmers to controlled shear flows in glass capillaries. Bending facilitated passage through highly curved microchannels. The deformation was elastic and swimmers completely recovered their shape after passing through the corner under the externally applied flow with a rate of 2 ml/min (Fig. 3*D*). Increasing the stiffness of the filaments reduced deformation and led to obstruction of the channel (Movie S5).

We conducted experiments in which the filaments were transported by a constant volume flux through a cylindrical channel with a constriction. Snapshots of experimental results are given in Fig. 3*E*. The passage through the constriction can be accommodated by a number of forces. Streamlines of the flow follow the conduit geometry, thus the hydrodynamic drag forces exerted on a flexible filament produce a deformation that facilitates passage. These forces have components along the conduit axis and along a normal axis pointing inwards, while both components promote passage. On one side, shear induced elongation due to the difference in flow rates experienced by different parts of the body as it passes though the constriction decreases the radius of the helix, thus facilitates passage. On the other side, the component of the hydrodynamic force pointing towards the conduit axis tends to compress the filament. This effect plays a more dominant role for helices going through sharp constrictions. The presence of the wall also modulates hydrodynamic forces acting on the filament. Lubrication stresses in the direction normal to the confining wall must be considered in the case of very narrow constrictions. In order to systematically explore the effect of shear-induced elongation, we built a numerical model by approximating the filament as a series of elastic segments of uniform helical shape (shown in the top panel of Fig. 3F). See *S5. Elastic helical filament passing through a constriction* for the details of the formulation. Snapshots of relevant experimental results and numerical simulations are given in Fig. 3*E* and Fig. 3*F*, respectively (see Movie S6). The simulations were based on the experimental conditions and measured value of the Young's modulus (E = 9 kPa). The helix has 3 turns, radius 1.25 mm, contour length 33.3 mm and helix angle $0.25\pi$ in its reference configuration.

As the helical filament entered the constriction, its axial length increased due to the higher flow rates experienced by parts of the helix closer to the constriction. This observation was faithfully captured in the simulation results. The plots of the speed of the two ends of the helix and the change in axial length were shown in Fig. S6A. The front-end speeds up as the filament enters the constriction, elongating the helix, and slows down as the filament exits the constriction, uniaxially compressing the helix. This leads to a deformed spiral shape, of reduced local radius towards the side that was further in the constriction. The decrease of radius accompanying the elongation, enabled the helix to pass through the channel. Reduction in the flow rate generated a mirrored shape at the exit of the constriction and the helix eventually regained its original shape (Fig. 3*E*). All these shape transformations were qualitatively captured by the simulations (Fig. 3*F*), which opens up the possibility to program the deformation of microswimmers for a given flow profile. Machines with a tubular body are less able to perform this accordion move as the tubular body cannot be stretched by the shear stress. At flow rates higher than 5 ml/min, all tested machines passed through constrictions smaller than their diameter simply by getting compressed between the walls of the channel. Navigation based on squeezing comes with the risk of obstructing the channel depending on the surface roughness and chemistry of the machines as well as the channel.

**Autonomous shape-shifting driven by osmolarity.** An appealing strategy to utilize different body plans for navigating in heterogeneous fluids is engineering a shape transformation triggered by the microrheology of the fluid. Incorporation of stimuli-responsive materials opens the door to fabrication of microdevices that can react to changes in ambient temperature, pH, or osmolarity. Bacterial movement in search of environments with optimal water content is termed osmotaxis. Experiments done with *E. coli* in polymer solutions revealed a long-term increase in swimming speed *(41)*, which scale with the osmotic shock magnitude *(42)*. Inspired by this mechanism, we tuned the mechanical and swelling properties of the nanocomposite by adding a hydrophilic co-monomer and reducing the cross-linking degree (*S6. Programmable shape transformation* and Movie S7). The increase in osmolarity dehydrates the swollen hydrogel and thereby reconfigures the body shape (Fig. S7).

Our data suggests that a tubular body with planar tail is preferable for swimming at low viscosity while a helical morphology would perform better at high viscosity. We built a reconfigurable microswimmer programmed to undergo a shape transformation between these two configurations in response to an increase in sucrose concentration (Fig. 4*A*). While the motility gradually decreased with increasing viscosity, the step-out frequency increased due to reduction of body size. Step-out frequency denotes the maximum rotating speed at which the swimmers can still synchronize with the external rotating magnetic field. Reduction of body size provided two enhancements that led to higher step-out frequency, higher magnetization and lower drag force. With the ability to increase the rotating frequency, the machines could be operated at a higher velocity. Therefore, the microswimmer with the programmed shape change exhibited a sustained velocity and enhanced maneuverability despite the increase in viscous forces (Fig. 4*B*). To our knowledge, this is the first time an artificial microswimmer increases its maximum rotating speed and maintain forward velocity with increasing viscosity. On the other hand, non-reconfigurable swimmers with the same initial configuration suffered from significant drop in motility (Fig. 1*F*) and longer reorientation time (Fig. 2*E*) at higher viscosities. Different polymorphic forms are observed in nature under changing solvent conditions such as pH value, salinity, and temperature. To investigate the propulsion provided by reconfigurable helices, we developed two types of configurations that respond to changes in osmolarity by continuously coiling (Type I) or uncoiling (Type II), respectively (Fig. 4*C* and Fig. S8). Type II helices sustained their motility despite the changes in viscosity while Type I helices were slowed down by increasing drag (Fig. 4*D*). On the other hand, Type I helices performed better when it came to navigation as expected.

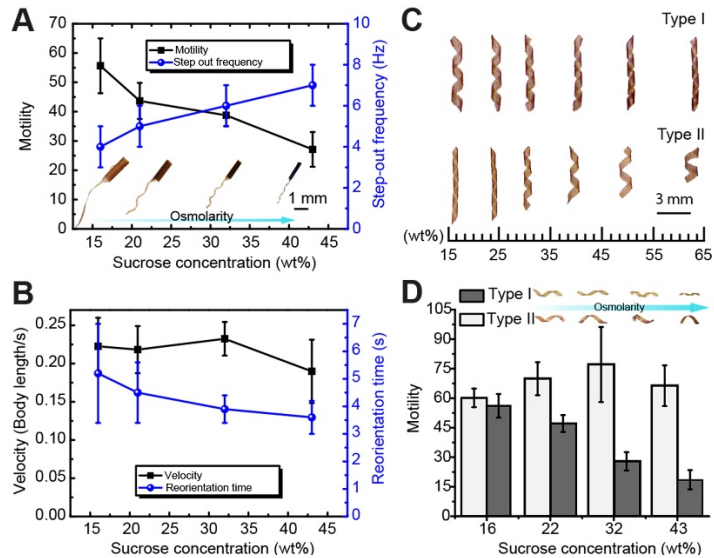

**Figure 4. Optimization of motility through shape shifting driven by osmotic or shear stress.** (*A*) The motility and step-out frequency of microswimmers in response to changes in sucrose concentration. (*B*) The sustained velocity and enhanced maneuverability of microswimmers in response to changes in sucrose concentration. The optical images show the effect of osmotic stress on the body and tail shapes. (*C*) Polymorphic transitions driven by osmolarity. Type I: continuously coiling with increasing osmolarity. Type II: continuously uncoiling with increasing osmolarity. (*D*) The motility of the microswimmers can be kept constant by utilizing polymorphic transitions to counteract viscous drag. All bar graphs represent average ±s.e.m. (N = 6 measurements for each microswimmer and three different swimmers tested per condition.

## Conclusion

In summary, we utilize magnetic hydrogel nanocomposites as a programmable matter to engineer microswimmers inspired by the form, locomotion, and plasticity of model microorganisms. We present methods for dynamic modulation of shapes, magnetization profiles, and locomotion gaits on the same device. A careful analysis of swimming performance at different viscosities provided a guideline to build a single machine that manifests multiple stable configurations, each optimized for a different locomotion gait. Shape adaptation in response to mechanical constraints and variation in osmotic pressure is performed via the coordination between the elastic and viscous stresses. Our approach for solving the navigation problem reduces the number of elements to be controlled and therefore can have advantages in terms of speed, versatility, and cost. The manufacturing process is high-throughput and scalable, which together open up doors for the development of a variety of adaptive soft microrobots.

## Methods

**Materials.** N-Isopropylacrylamide (NIPAAm) as the monomer, Acrylamide (AAm) as the hydrophilic co-monomer, 2, 2-dimethoxy-2 phenyla-cetophenone (99%, DMPA) as the photo-initiator, and Poly(ethylene glycol) diacrylate (average MW 575, PEGDA) as the cross-linker were all purchased from Sigma-Aldrich. Magnetic nanoparticles (Dispersible 1% polyvinylpyrrolidone (PVP) coated 30 nm magnetite) was obtained from Nanostructured and Amorphous (USA).

**Fabrication of monolayer machines.** Detailed fabrication processes of the monolayer structures can be found in our previous work *(27)*. A pre-gel nanocomposite solution (NIPAAm–AAm–PEDGA and MNPs) was injected into a chamber constructed by sandwiching the glass mask and a silicon wafer with patterned SU-8 spacers. The SU-8 spaces define the thickness of the chamber, which is 30μm. For the non-reconfigurable microswimmers, the molar ratio between NIPAAm-AAm-PEGDA was 100-0-5, and the concentration of MNPs was 5 wt%. Structures with smaller folding radius was obtained by increasing the MNP concentration to 10 wt%. After the pre-gel solution filled the entire chamber, a gradient of MNPs was generated through gravitational sedimentation. Subsequently, a static uniform magnetic field with a strength of 10 mT and ultraviolet exposure (365 nm, 3 mWcm$^{-2}$) were simultaneously applied to the pre-gel solution in order to align the MNPs while polymerizing the pre-gel solution (Fig. S2*A*). After curing, the sandwich construction was opened. The structures were released from the glass mask by submerging into water.

**Experimental platform.** The 3D alignment of the MNPs was performed using a solenoid in horizontal direction with an inner diameter of 5.5 inch in conjunction with a pair of Helmholtz coils separated by 5.5 inch in vertical direction (Fig. S2A). Ultraviolet lamps (Lightening Enterprises, USA) were integrated inside the solenoid to initiate the cross-linking of hydrogel polymer. The maximum strength of generated uniform magnetic fields by the solenoid and Helmholtz coils are 10 and 5 mT at the center region, respectively. The motion studies were conducted with a custom-design eight-coil electromagnetic manipulation system (Fig. S2C), which is called Octomag *(43)*. The maximum uniform magnetic field generated by the system is 40 mT, and the maximum magnetic field gradient is 1 T/m. We produced the confined channel employed for hydrodynamic uncoiling experiments by connecting two glass pipettes with a glue gun. One of the open ends of the confined channel is connected to a syringe as shown in Fig. S2B. A syringe pump controls the flow rate. The viscosity of the sucrose solution was measured using TA Instruments Rheometer AR 2000. Please see the supporting text for the details of characterization.

**Preparation of osmolarity-responsive microswimmers.** For the fabrication of Type I reconfigurable microswimmers a hydrophilic monomer Acrylamide (AAm) was incorporated into the gel solution with a molar ratio (NIPAAm: AAm: PEGDA) of 85:15:1. The molar ratio of Type II is set as 100:0:1.

**Acknowledgments**

**General**: We thank Alexandre Persat, John McKinney, Pedro Reis, and Ece Ozelci for fruitful discussions; Chen-Pu Hsu, Wenjie Sun, Xiangzhong Chen and Carlos Alcantara for assistance with experiments; and Qianwen Chao for assistance with data analysis. All data needed to evaluate the conclusions in the paper are present in the paper and/or the Supplementary Materials. Additional data related to this paper may be requested from the authors.

**Funding:** This work was supported by the European Research Council under the ERC grant agreements ROBOCHIP (714609), PhyMeBa (682754) and SOMBOT (743217).

**Author contributions:** H.-W.H., M.S.S., and B.J.N. designed research; H.-W.H. and M.S.S. performed research and analyzed data; F.E.U. developed the analytical theory; P.K. and E.L. formulated and implemented the computational model; H.-W.H. and M.S.S. wrote the paper with contributions from all authors.

**Competing interests:** The authors declare no conflict of interest.

**Supplementary Material**

**S1 Programmable folding of monolayer nanocomposites.** The immediate consequence of incorporating MNPs into hydrogels is the change in their weight-swelling ratio (WSR). The WSR of a hydrogel determines the expansion potential of the material, and it is primarily determined by the material composition. Different concentration of the cross-linker and MNPs in the hydrogel nanocomposite are tuned to attain various WSR (Fig. S1*A*). Increasing the concentration of cross-linker (X) and MNPs monotonically decreases the WSR stemmed from the increased cross-linking degree in the hydrogel and enhanced stiffness by adding MNPs. Thereupon, the difference in WSR of the hydrogel was acquired before and after adding MNPs with varying X (Fig. S1B). The highest difference in WSR is given by X = 1. While, X = 0.5 indicates the lowest difference in WSR in the wake of its comparatively high swelling property minimizing the interaction between the MNPs and the polymer network. The mechanical properties of the hydrogel comprising elastic moduli and expansion coefficient are primarily associated with WSR. Higher WSR results in larger expansion coefficient but weaker mechanical strength.

The gradients of magnetic nanoparticles generated by gravitational sedimentation was sufficient to generate folding of monolayers. The gradient of MNPs induces a gradient of swelling ratio along the thickness, resulting in a bending of the thin sheet from top to bottom upon swelling and contracting. Assuming that the arc length is proportional to the expansion of the thin film (Fig. S1*C*), the folding curvature of monolayer thin films can be approximated as:

$$\frac{\left(R+\frac{1}{2}\right)\cdot\vartheta}{\left(R-\frac{1}{2}\right)\cdot\vartheta} = \frac{\alpha_1 \cdot l}{\alpha_2 \cdot l} \tag{S1}$$

The curvature of the self-folded monolayer composite is given as follows:

$$K = \frac{1}{R} = \frac{2}{t}\left(\frac{\eta}{2-\eta}\right) \tag{S2}$$

where $\eta = (\alpha_1 - \alpha_2)/\alpha_1$, $\alpha_1$ and $\alpha_2$ is the expansion coefficient of the upper part and bottom part of the nanocomposite, which are both function of WSR and concentration of MNPs. Equation (S2) indicates that the higher the difference in the expansion coefficient between the top and bottom layers the larger the folding curvature is.

To characterize the minimum time required to complete sedimentation, we waited for varying durations of time before polymerizing the nanocompoite. Experiments showed that one hour is sufficiently long to obtain repeatable folding and the curvature does not change markedly with longer waiting time (Fig. S1*D*). Increasing the concentration of MNPs increases the curvature of monolayer structure as expected (Fig. S1*E*). However, at concentrations higher than 10 wt%, the reflection from nanoparticles became significant, which resulted in an incomplete photopolymerization. To address this problem, the thickness of the nanocomposite layer was changed. Indeed, by reducing the thickness from 30 μm to 10 μm, the nanocomposite layers with higher nanoparticle concentrations (20 wt%) could be polymerized. Additionally, the WSR of nanocomposites was modified through tuning the cross-linker ratio and tested if increasing the WSR would monotonically increase curvature. While the curvature of helices increased in response to an increase in WSR from 3 to 4, further increase did not lead to a significant change until reaching a threshold. For WSR values above 10, the swelling ratio of the hydrogel became comparable to the swelling of the nanocomposite (Fig. S1*F*). Thus, the overwhelming status of particle gradients in programming folding tend to be faded away, and the curvature of helices started to reduce in the wake of the approximately homogenous expansion.

**S2 Body size effect on flagellated microswimmers.** Figure S3 shows a very simple model of a flagellated microswimmer in compression, consisting of two rigid, slender rods connected by a torsional spring (hook) of constant modulus. A propulsive force $F_P$ exerted by the helical tail pushes the machine body. The front end of the body is pinned at the origin while free to rotate. The axis of the body and the tail are in parallel with an offset distance of the radius ($R$) of the body. This offset tends to promote a buckling at the connection of the body and the tail, which in turn results in precession of the body manifesting a precession angle $\theta_P$. The angular velocity of the bending angle is given by

$$\dot{\theta}_P = \frac{1}{C_\perp} \sum T_{ext} \approx F_P \cdot R, \tag{S3}$$

where $C_\perp$ is the rotational resistance normal to the plane of the machine body and $T_{ext}$ is external drag torques subjected on the machine body. Equation (S3) suggests that increasing the radius of the machine gives rise to increase the precession angle.

There is an alternative way of building a similar argument. If we approximate the rotating body as a rod spinning around its long axis, the viscous drag is given by *(44)*:

$$\tau_d = \left(4\pi \mu_d L_B \omega r_f^2\right)/\sqrt{1-k^2} \tag{S4}$$

where $\mu_d$ is the viscosity of the fluid, $\omega$ is the rotational speed, $r_f$ is the folding radius of the tubular body, $L_B$ is the length of the tubular body, $k = r_f/d$, and $d$ is the distance between the center of the microswimmer and the wall. Equation (S4) suggests that reducing the radius of the body will considerably decrease the rotational friction and lead to a decrease in the body precession.

**S3 Characterization of the magnetization profile.** We characterized the magnetic properties of thin film magnetic nanocomposites using vibrating sample magnetometry (VSM). A low magnitude magnetic field is applied with values ranging from -10 mT to 10 mT. The same level of magnetic field was applied for reinforcing the MNPs during photopolymerization. Disc-shaped thin-film nanocomposites (DTNs) are chosen due to the in-plane shape anisotropy. That is to say, DTN can be magnetized in any direction of the disk plane if the embedded MNPs are uniformly dispersed. We applied a magnetic field orthogonal to the alignment of MNPs and measure the magnetic moment along and orthogonal to the alignment (Fig. S4*A*). We observed a strong and permanent magnetic moment along the alignment of MNPs. If DTNs were free to move, they would spontaneously align the dipole moment to the external magnetic field.

The encoded permanent dipole moment created by the alignment of MNPs can be re-coded by applying a stronger magnetic field in any desired direction other than the alignment of MNPs. To demonstrate this recoding magnetization, we applied a magnetic field ranging from -30 mT to 30 mT and measure the resulting magnetic moment (Fig. S4*B*). The permanent moment created by the alignment of MNPs was largely attenuated by the stronger magnetic field and became smaller than the magnetic moment along the applied magnetic field. These results indicate that the encoded structure was re-magnetized in the direction orthogonal to the alignment of MNPs.

On that basis, Fig. S4*C* and S4*D* show the frequency response wobbling angle of a helical microswimmer before and after recoding its magnetization profile, respectively by means of the proposed approach of re-magnetization. Microswimmers with $0 < \phi < 90°$ wobble around their long axis at low frequencies. With increasing frequency, the wobbling angle gradually decreases, and eventually leads to a corkscrew motion (Fig. S4*C*). When the amplitude of the field was increased from 20 mT to 30 mT, a new magnetization profile was encoded during swimming where $\phi = 90°$. After successful reprogramming of magnetization profile, swimmers performed a wobble-free corkscrew motion at lower rotation frequency (Movie S3). This strategy allows to dynamically re-magnetize microswimmers and adjust $\phi$ at any angle from $0°$ to $90°$ by applying a strong rotating magnetic field with a conical angle corresponding to the desired $\phi$.

**S4 Analytical solution for an actuated elastic filament in a viscous fluid.** To predict the shape of the elastic tail during swimming, elastic and hydrodynamic forces must be solved together. Elastic forces on the slender filament, given in equation (S5-15), are coming from the bending energy and the inextensibility constraint *(45)*.

$$\varepsilon = \int_0^L \left[\frac{A}{2}\kappa^2 + \frac{\Lambda}{2}r_s^2\right] ds, \tag{S5}$$

where $\kappa$ is the curvature of the tail ( $\kappa = \psi_s$) and $\Lambda$ is the Lagrange multiplier. We can derive the elastic force per unit length of the filament, by taking $f_\varepsilon = \partial \varepsilon / \partial r$.

$$f_\varepsilon = -(A\psi_{sss} - \psi_s \tau)\hat{n} + (A\psi_{ss}\psi_s + \tau_s)\hat{t}. \tag{S6}$$

where $\hat{n}$ is unit normal and $\hat{t}$ is unit tangent vectors. The local tension in the filament is denoted by $\tau_s$. Resistive Force Theory (RFT) is used to model the hydrodynamics of the shape adaptation phenomenon. RFT assigns values to local drag coefficients and these coefficients relate local viscous forces per unit length to local parallel and perpendicular velocities *(46)*.

$$\xi_\perp = \frac{4\pi\mu_d}{ln\left(\frac{L}{r}\right)+0.5}; \xi_\parallel = \frac{2\pi\mu_d}{ln\left(\frac{L}{r}\right)-0.5}. \tag{S7}$$

Velocity vector of an infinitesimal part is found through the time derivative of position vector of part $r_t$. When we define the fluid velocity in the fixed frame as **u**, total velocity is **u** – $r_t$. The drag force per unit length of the filament is given by:

$$f_d = -[\xi_\perp \hat{n}\hat{n} + \xi_\parallel \hat{t}\hat{t}] \cdot (r_t - u), \tag{S8}$$

When local mechanical equilibrium between elastic forces (S6) and hydrodynamic forces (S8) is achieved, partially differential equation becomes an hyperdiffusion equation given by:

$$\frac{\partial y}{\partial t} \simeq -\frac{A}{\xi_\perp}\frac{\partial^4 y}{\partial x^4} \tag{S9}$$

Non-dimensionalization of equation (S9) gives us Sperm number, $S_p$. We assume that one end of the filament is held fixed (x = 0) and oscillated vertically in a sinusoidal form with an amplitude $\epsilon$ ($y|_{x=0} = \epsilon \cos(wt)$). This end is free of bending moment since it is hinged ($y_{xx}|_{x=0} = 0$). The other end of the filament (x = 1) is free to move, therefore there is no force or bending moment, which yields $y_{xxx}|_{x=1} = 0$ and $y_{xx}|_{x=1} = 0$, respectively.

Motion of the filament is confined in the xy plane. Using the linearity condition at x = 0 allows to use a harmonic solution in time in the form of

$$y = \epsilon\ Re[e^{it}h(x)] \qquad (S10)$$

Equation (S9) is transformed into an ordinary differential equation in the form of

$$ih = -Sp^{-4}\frac{d^4h}{dx^4} \qquad (S11)$$

Solution for Eq. (S11) must be in the form of $h = ce^{kx}$, where c is a constant and k is determined as:

$$k^4 = iSp^4$$
$$k_n = i^n e^{-i\pi/8} Sp, \text{ n = 1,2,3,4} \qquad (S12)$$

When all the modes are superimposed, a general solution for $h$ can be found as:

$$h(x) = \sum_{n=1}^{4} c_n e^{k_n x} \qquad (S13)$$

where $c_n$ are derived by using the following boundary conditions:

$$\sum_{n=1}^{4} c_n = 1,$$
$$\sum_{n=1}^{4} k_n c_n = 0,$$
$$\sum_{n=1}^{4} k_n^2 e^{k_n} c_n = 0,$$
$$\sum_{n=1}^{4} k_n^3 e^{k_n} c_n = 0. \qquad (S14)$$

Finally, deformation of the filament is given with the following time-dependent equation:

$$y(x,t) = \epsilon Re[\ e^{it}\ h(x)] = \epsilon Re[\ \sum_{n=1}^{4} c_n e^{k_n x + it}] \qquad (S15)$$

We solved equation (S15) using parameters extracted from the experiments presented in Fig. 3B. Length, thickness and width of the filament were taken as 2 mm, 30 µm, and 100 µm, respectively. Elastic modulus of the filament was taken as 10 kPa. Fig. S5 shows the simulation results for three different cases corresponding to the experiments presented in Fig. 3B. The coiling behavior observed with increasing viscosity and frequency was successfully replicated in the solution of the analytical equations.

**S5 Elastic helical filament passing through a constriction.** To explore the shear-induced elongation effect that facilitates passage of a helical filament through a constriction, we followed a semi-analytical approach. We consider an elastic filament, of helical shape in its reference, unstressed configuration, with angle $\alpha_0$, radius $R_0$ and total contour length $L$ as shown in Fig 3F, that is convected by a flow driven by a constant volume flux $Q$ (by mass conservation), through a cylindrical channel aligned along the z-axis, with a slowly-varying, axisymmetric constriction of cross-sectional radius $a(z)$

$$a(z) = \alpha_0(1 + \epsilon \tanh^2(z/z_c)) \qquad (S16)$$

Our model approximates the filament as a series of short, uniform helical segments, that are defined by labelling $N$ material points at contour-length positions $s_i^{(0)} = (i-1)\Delta s_0$, $i = 1, \cdots, N$, $N$ and axial positions $z_i^{(0)} = s_i^{(0)} \cos \alpha_0$ in the reference configuration, as shown in Fig. 3$E$ and Fig. S6$B$. We assume each of these $N$-1 small segments of the filament between neighboring material points will always remain a uniform helix of axial length $z_{i+1}(t) - z_i(t)$, constant contour length $\Delta s_0$ and helix angle $\alpha_i$, given by the inextensibility condition, the requirement that the contour length of each segment is conserved,

$$\cos \alpha_i = \frac{z_{i+1} - z_i}{z_{i+1}^{(0)} - z_i^{(0)}} \cos \alpha_0 \tag{S17}$$

Assuming the spheres of radius $r$ at each material point are small enough so as not to affect the flow, the force balance on the $i^{th}$ sphere is

$$-\xi r [\dot{z}_i(t) - U(z_i(t))] + f_{el}(\zeta_{(i+1),i}) - f_{el}(\zeta_{i,(i-1)}) = 0, \tag{S18}$$

where $\xi = 6\pi \mu_d$, $i \in \{2, \cdots, N-1\}$. The first term is the viscous drag on the sphere of radius $r$ attached at the material point $i$ moving at speed $\dot{z}_i(t)$ in the background flow $U(z_i(t)) = Q/(\pi a^2(z))$, the second and the last terms are the elastic forces from the elastic segment between ($i$+1,$i$) and ($i$,$i$-1), respectively. The elastic forces can be expressed as *(47)*

$$f_{el}(\zeta) = -\frac{EI \sin^2(\alpha_0)}{R_0^2} \sqrt{1-\zeta^2} \cdot \left(\sin(\alpha_0) + \frac{h\zeta \cos(\alpha_0)}{\sqrt{1-\zeta^2}}\right)\left(h\cos(\alpha_0) - \frac{\zeta \sin(\alpha_0)}{\sqrt{1-\zeta^2}}\right), \tag{S19}$$

Equation S19 gives the elastic force arising from the axial extension of a helix to a normalized (by the contour length) axial length $\zeta = \Delta z/\Delta s_0$. We use the notation $E$ for the Young's modulus and the expression $I = \frac{\pi r^4}{4}$ for the principal moment of inertia for a filament with a cylindrical cross-section of radius $r$.

In equation (S18), $\zeta_{(i+1,i)} = (z_{i+1}(t) - z_i(t))/\Delta s_0$ is fractional axial length of the small helical segment between the material points at $z_i$, $z_{i+1}$, and similarly for $\zeta_{(i,i-1)} = (z_i(t) - z_{i-1}(t))/\Delta s_0$. For the first and last material points at the two endpoints of the neighboring filament, the force balance is similar, except that one of the filaments is missing in each case.

For convenience, we take the material points at the ends of each helical turn, and assume the helical segment between them to always have one turn. Then the radius $R_i$ of the helical segment between points, $z_i, z_{i+1}$, is given by

$$R_i = \frac{\sin(\alpha_i)}{\sin \alpha_0} R_0. \tag{S20}$$

In order to produce snapshots of the simulation in the x-z plane, we plot each segment as

$$x_i(z) = R_i \sin\left(\frac{\sin(\alpha_0)}{R_0 \cos \alpha_i}(z - z_i)\right), \quad z \in (z_i, z_{i+1}). \tag{S21}$$

Fig. S6 shows the shape of the constriction, the time evolution of the endpoint speeds and the axial length of the resulting spiral, for the simulation whose snapshots are shown in Fig. 3F. The simulation considers an elastic helix with elastic Young's modulus E = 9 kPA with 3 turns, radius 1.25 mm, contour length 33.3 mm and helix angle π/4 in its reference configuration. Convected by a flow of volume flux 5 mL/min, the helix passes through a constriction that narrows from a radius of 1.6 mm to the narrowest radius of 1.3 mm and characteristic width of $z_c$ = 33.3 mm. The value of ε in Eq. S16 is taken to be 0.3, and the value for the cross-sectional radius $r$ as 50 μm. All these parameters were taken from experimental measurements. As the helical filament enters the constriction, its axial length increases, as shown in the relevant plot of Fig.S6A , due to the higher flow rates experienced by parts of the helix closer to the constriction as shown in the plot of the endpoint speed. This leads to a deformed spiral shape, of reduced local radius towards the side that is further in the constriction. The decrease of radius enables the helix to pass through the channel. As the helix exits the constriction, it is axially compressed, due to the higher flow rate experienced by the back parts of the filament that are closer to the constriction compared to the front parts, as shown in Fig. S6A.

**S6 Programmable shape transformation of monolayer structures.** The reconfigurable microswimmers were attained by reducing the cross-linking degree of the hydrogel. Fig. S7*A* shows the WSR of the hydrogel with X = 5 and X = 1 with varying temperature. The hydrogel with X = 5 is inert to temperature variation, which makes the microswimmers non-reconfigurable. While, the hydrogel with X=1 shows remarkable variation in WSR with varying temperature, which imparts reconfigurable shapes to the microswimmers. Fig. S7*B* shows the mechanical properties of the hydrogel with different WSR.

Thus far, the shape transformation of NIPAAm hydrogel structures has been attained by dehydrating the polymer network through increases in ambient temperature. An alternative to dehydrating the hydrogel is to increase the osmolarity of the aquatic environments. In this work, we analogized the osmotic-responsive shape transformations to the temperature-responsive shape transformations, and studied the shape transformation of monolayer structures by varying the temperature.

To investigate the osmotic-response of the monolayer microswimmers, the monolayer structure is modeled as a bilayer comprising two nanocomposites with different concentrations of MNPs (2.5wt% vs 7.5wt%). The WSR of the two nanocomposites were characterized in the case of varying temperature as shown in the inset of Fig. S8*A*. The WSR of both nanocomposites decreased with increasing temperature and the layer with lower concentration of MNPs always had higher WSR. The difference in the WSR between the two nanocomposites can be calculated as

$$\eta = \frac{WSR_h - WSR_l}{WSR_h} \tag{S22}$$

where $WSR_h$ and $WSR_l$ denote the WSR of the nanocomposite with higher and lower concentration of MNPs, respectively. The difference in WSR corresponds to the $\eta$ of Eq. (S2).

Figure S8 shows that the difference in WSR first increased with the rising of temperature prior to 32℃ and thereupon continuously decreased with the rising of the temperature. The folding curvature of the monolayer structure increases with the rising of $\eta$, resulting in the coiling of monolayer helical configuration (Figure S8*B*). The decrease of $\eta$ resulted in the uncoiling of the helix. The transition temperature of the coiling and uncoiling were attained to be corresponding to the LCST of the hydrogel. Accordingly, the question is raised whether the self-folded helix can continuously coil in the case of increasing temperature prior to 45℃ by shifting the LCST. To test our hypothesis, helical-shaped monolayer structure that comprises NIPAAm, AAm and PEGDA with the molar ratio of 85:15:1 (Type II reconfigurable helix) was fabricated, resulting in the LCST of 45℃ (Fig. S8*C*). Fig. 4*C* shows that the osmotic-responsive shape transformations of type I and type II reconfigurable helices resemble the thermo-responsive shape transformation. The osmotic response of the sucrose concentration ranging from 16 wt% to 64 wt% is corresponding to the temperature response ranging from 32℃ to 50℃.

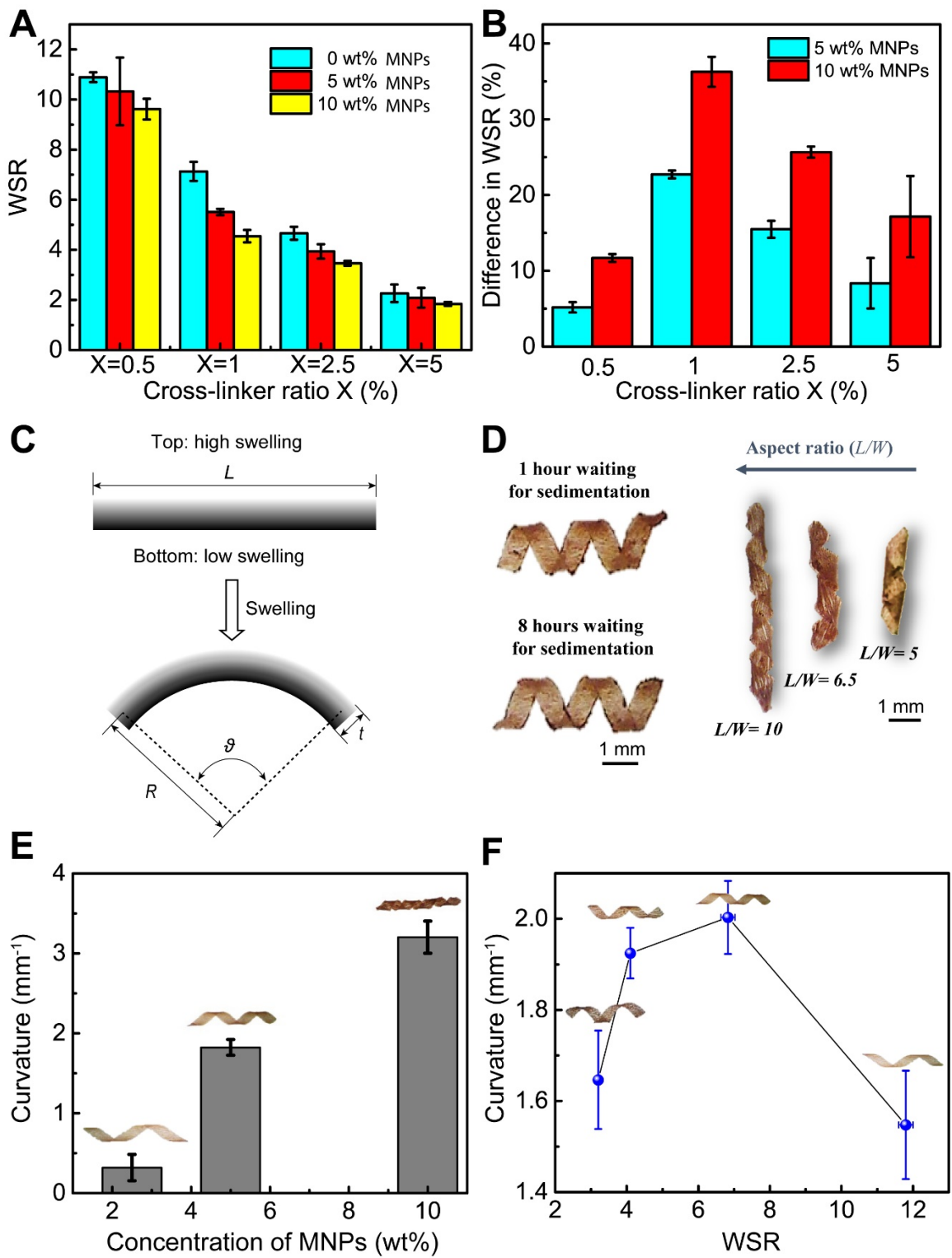

**Fig. S1.** Characterization of the folding mechanism of monolayer structures. (*A*) The effect of cross-linking degree in the weight swelling property (WSR) of the nanocomposites with various concentration of MNPs. (*B*) The difference in WSR between the hydrogel and the hydrogel nanocomposites with varying the cross-linker concentration. (*C*) Schematic illustration of self-folding monolayer upon hydration. (*D*) Gravitational sedimentation is finalized within one hour and waiting longer does not affect folded shape. Optical images of the helices folded from 2D patterns with different aspect ratio. Folding curvature of monolayer structures with varying (*E*) WSR and (*F*) concentration of the embedded MNPs.

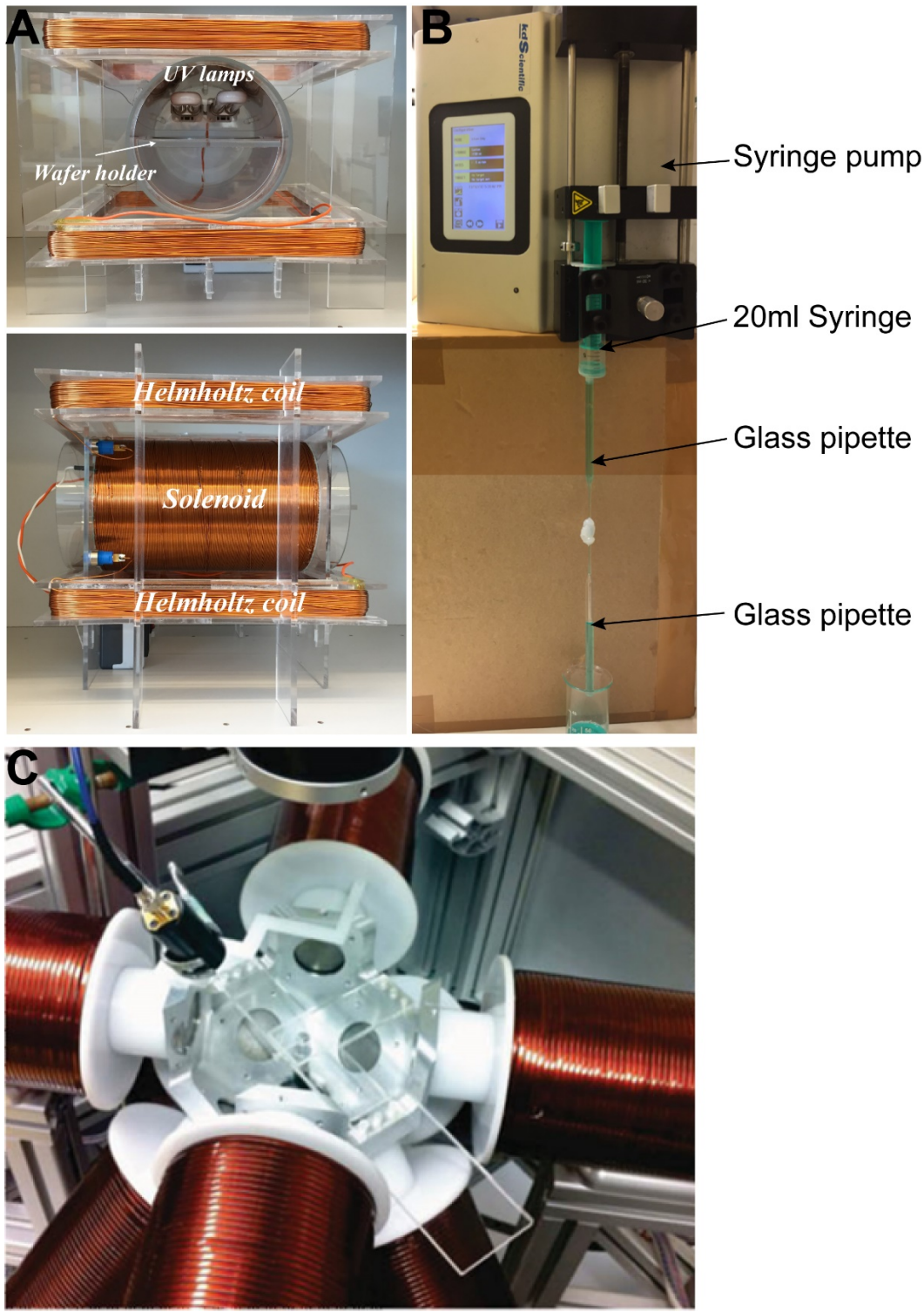

**Fig. S2.** Experimental platform. (*A*) Electromagnetic coil setup integrated with a UV light source for aligning MNPs during photopolymerization. (*B*) Setup for flow experiments include a constricted channel generated by connecting two glass pipettes and a plastic syringe connected controlled by a syringe pump. (*C*) Octomag electromagnetic manipulation system is used for actuating microswimmers.

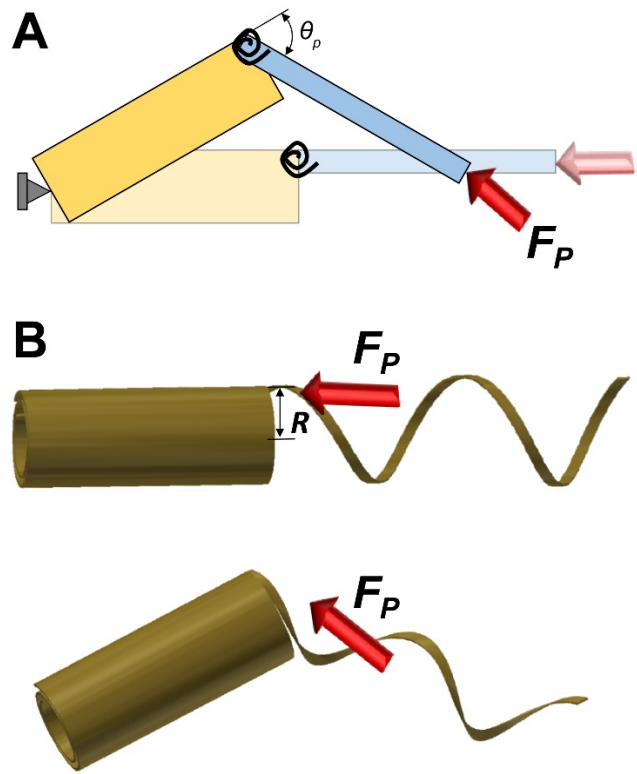

**Fig. S3.** (*A*) Static toy model showing the body and tail as two connected rods with different width. Propulsion is aligned with the tail. (*B*) Dynamic model shows a cylindrical body and helical tail. The propulsion is aligned with the tail.

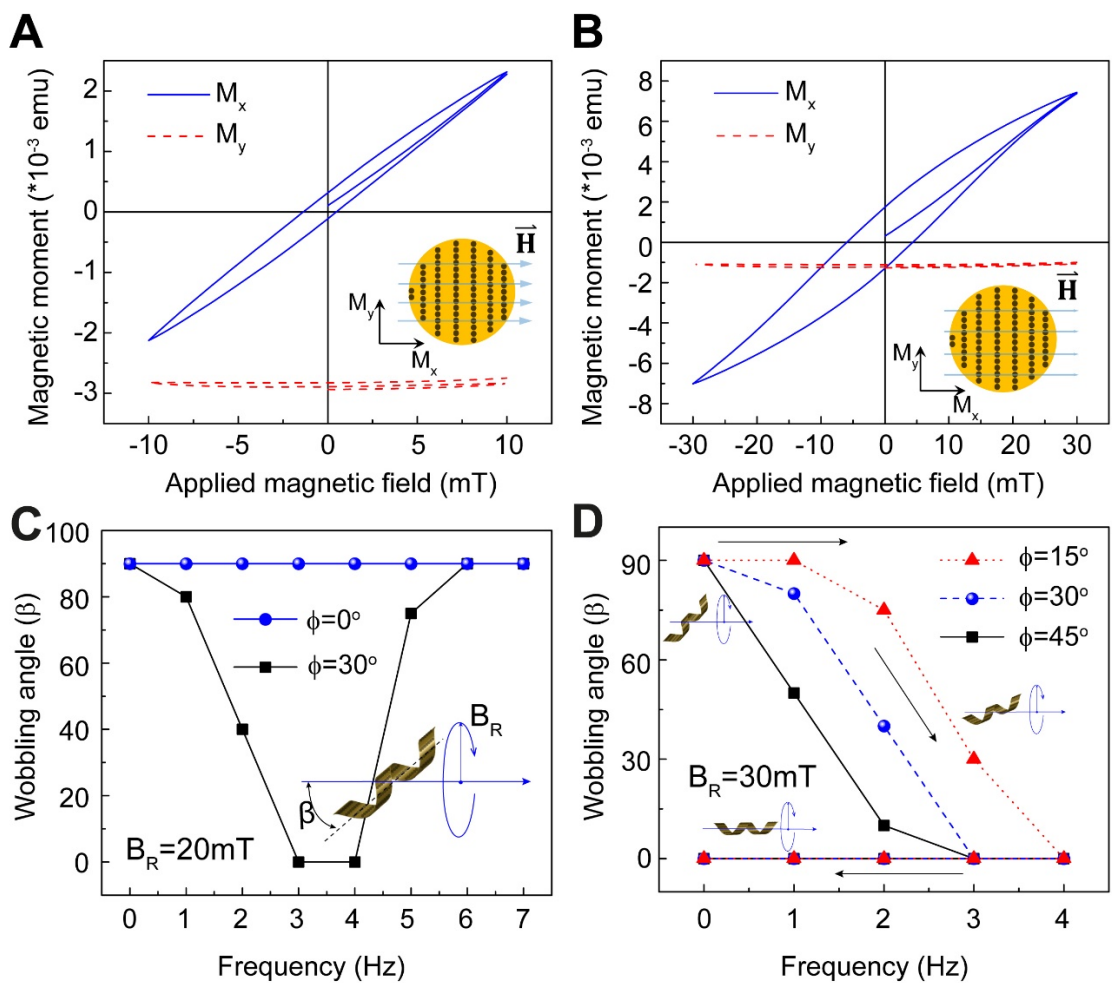

**Fig. S4.** Characterization of the magnetic properties of the programmable nanocomposites using vibrating sample magnetometry (VSM) measurements. Hysteresis loops of an in-plane reinforced nanocomposite generated by applying a magnetic field orthogonal to the alignment of MNPs with an amplitude of (*A*) -10 mT to 10 mT and (*B*) -30 mT to 30 mT. Wobbling angle of helical microswimmers with various misalignment angle $\phi$ versus frequency driven by rotation magnetic fields with intensity of (*C*) 20 mT and (*D*) 30 mT.

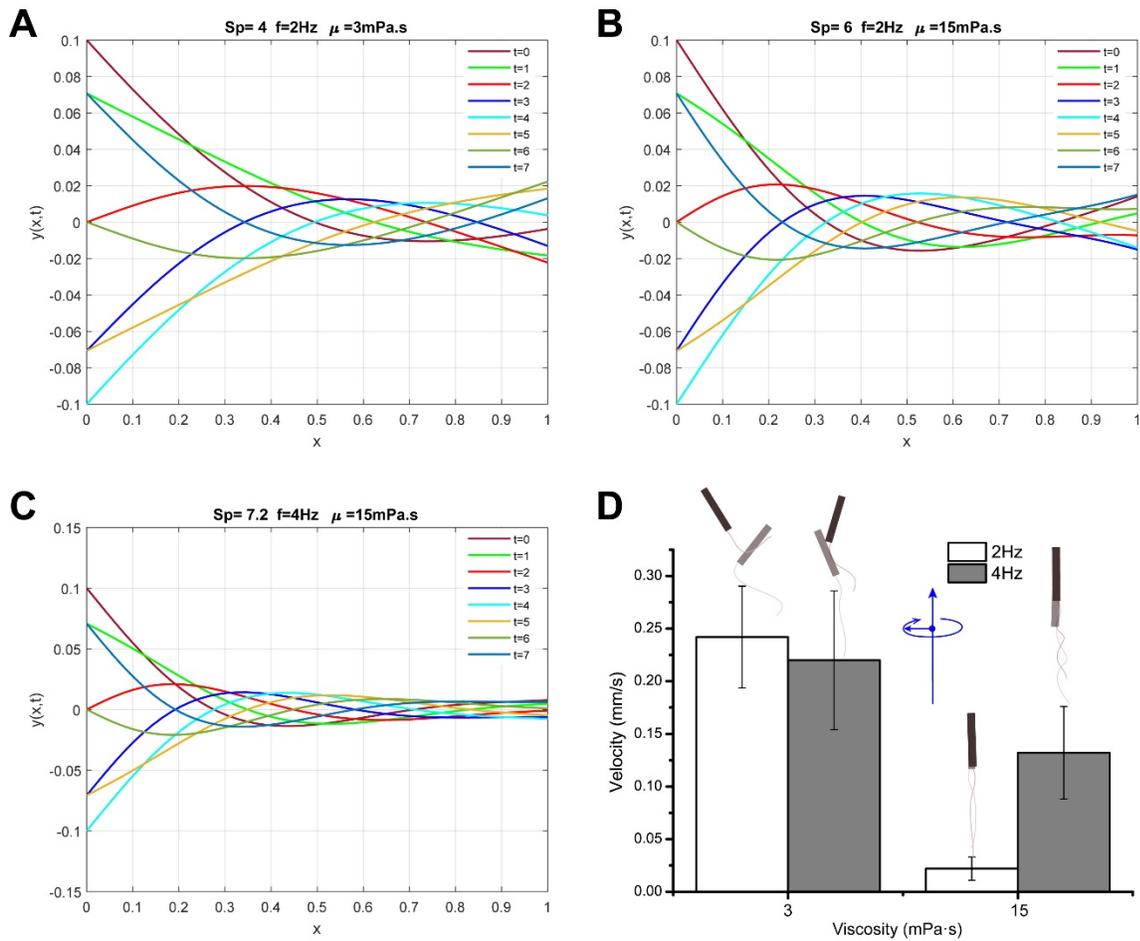

**Fig. S5.** Deformation of a flexible filament actuated from one end. The actuation is modeled as oscillations generated at one end of the filament while keeping the other end free to move. Different colors show the shape of the filament at seven different time points during one full cycle. (*A*) The deformation of the filament for Sp=4. The viscosity and frequency of the oscillation are given by 3 mPa.s and 2 Hz, respectively. (*B*) The deformation of the filament for Sp=6. The viscosity and frequency of the oscillation are given by 15 mPa.s and 2 Hz, respectively. (*C*) The deformation of the filament for Sp=7.2. The viscosity and frequency of the oscillation are given by 15 mPa.s and 4 Hz, respectively. (*D*) The velocity along with schematic representations of microswimmers with shape shifting tails moving at varying viscosities and rotating frequencies.

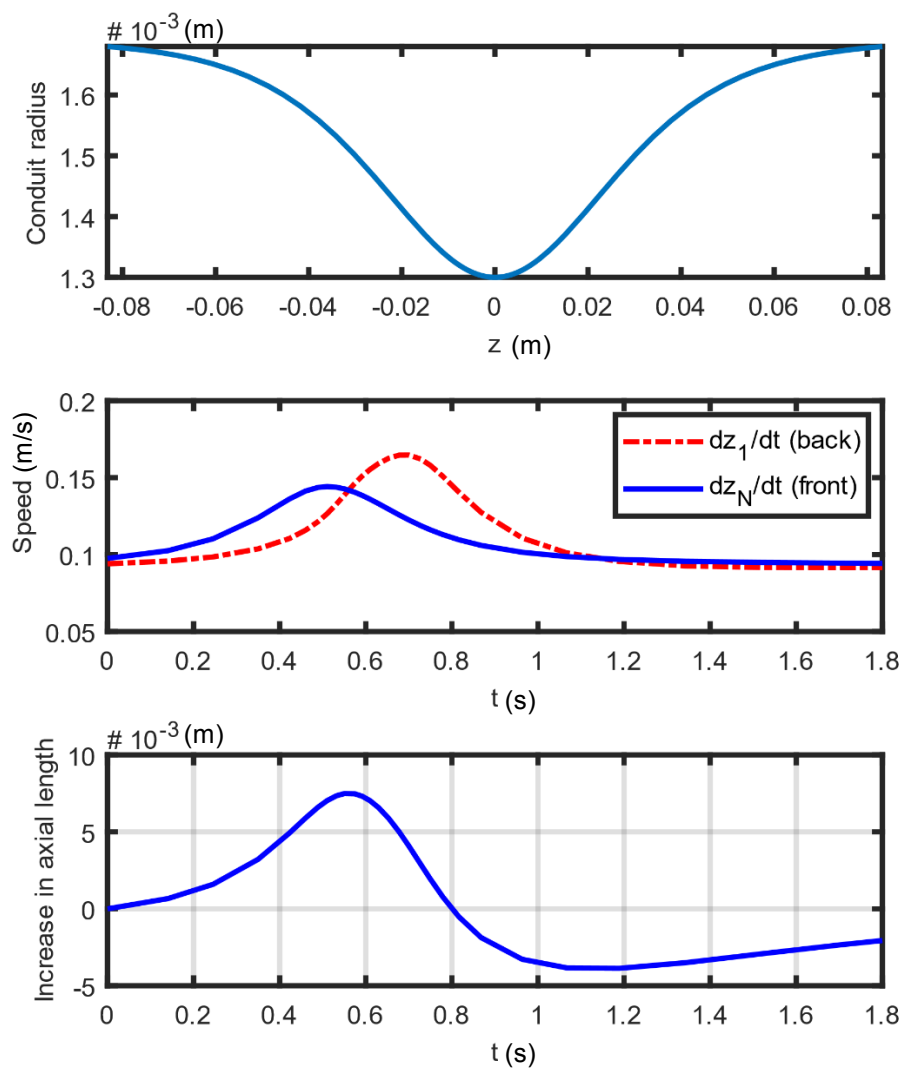

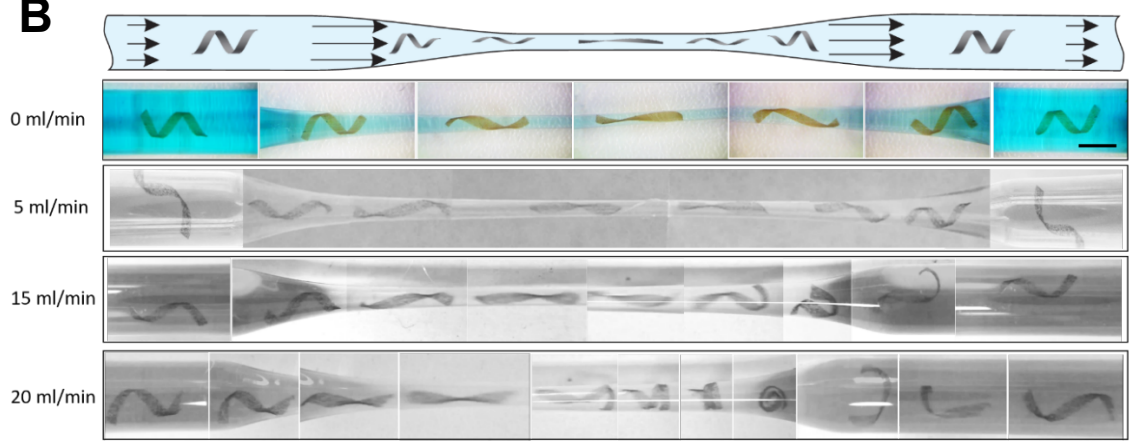

**Fig. S6.** (*A*) Plots of the radius variation along the constriction, the time evolution of the endpoint speeds, the axial length and the maximal radius of the resulting spiral. (*B*) The optical images of the flexible helix passing through the mechanical constraint at different flow rates.

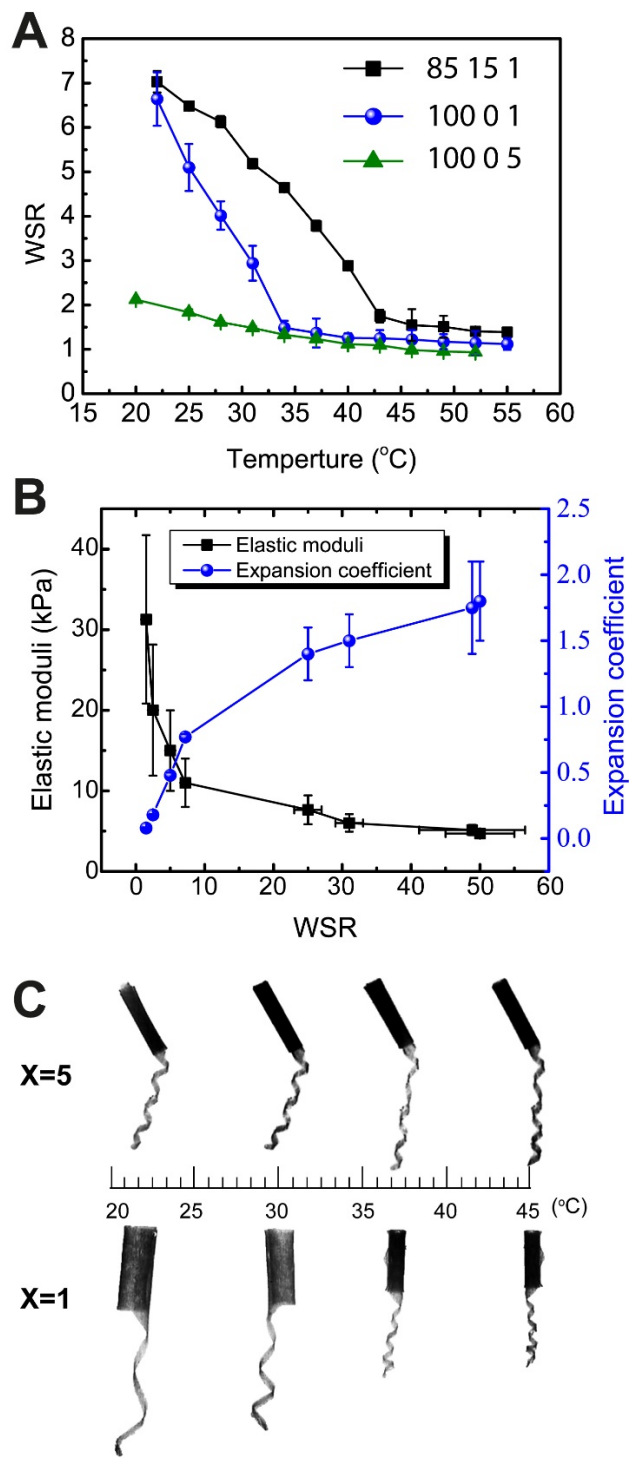

**Fig. S7.** Characterization of the mechanical properties of hydrogel with various swelling properties. (*A*) Temperature response of weight swelling ratio (WSR) of hydrogels with different ratio of NIPAAm-AAm-PEGDA. (*B*) Elastic moduli and expansion coefficient of the hydrogel versus WSR. The elastic modulus of the hydrogels with different cross-linking degrees were measured by uniaxial tensile test tool (242 Actuator, MTS system corp, Eden Praire, MN, USA) by using single layer sheets (50x10x2.5 mm) (*C*) The morphologies of the swimmers with different material compositions (NIPAAm-AAm-PEGDA: 100-0-X) at 22℃, 30℃, 37℃, and 45℃.

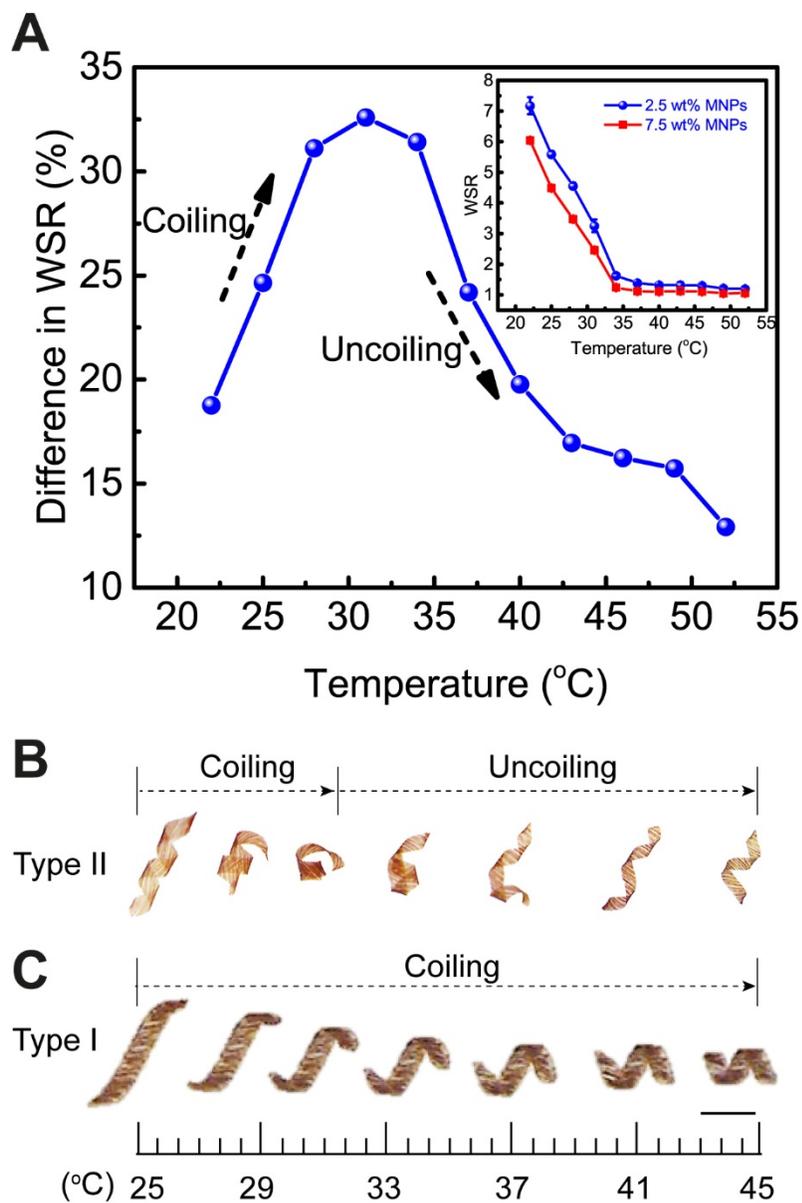

**Fig. S8.** The coiling and uncoiling mechanism of self-folding monolayer structure. (*A*) The difference in weight swelling ratio (WSR) of the hydrogels embedded with different concentration of MNPs versus temperature. (*B*) Type II monolayer structures with the molar ratio of NIPAAm-AAm-PEGDA at 100:0:1 experiencing coiling and then uncoiling while increasing temperature to be dehydrated. (*C*) Type I monolayer structure with the molar ratio of NIPAAm-AAm-PEGDA at 85:15:1 continuously coiling while increasing temperature to be dehydrated.

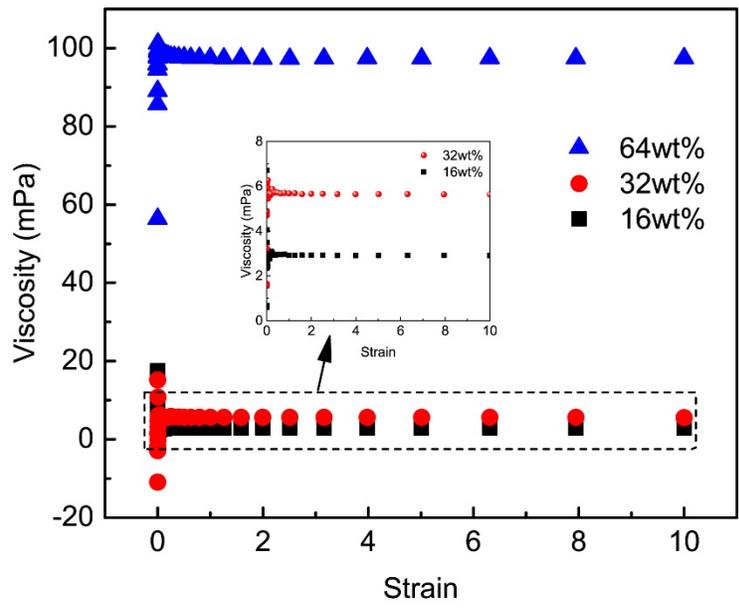

**Fig. S9.** The measured viscosity of the media with varying the concentration of sucrose.